\documentclass[12pt]{iopart}

\usepackage{iopams} 
 
\def\Pop#1#2{P^{(#1)}_{#2}}
\def\Xop#1#2{X^{(#1)}_{#2}}
\def\Xbop#1#2{\bar{X}^{(#1)}_{#2}}
\def\Jop#1#2{J^{(#1)}_{#2}}
\def\hf{\frac{1}{2}}
\def\JHEP{\textit{J. High Energy Phys.}\ }

\begin{document}

\title[]{$ {\cal N} = 2 $ Galilean superconformal algebras with central extension}

\author{N Aizawa}

\address{Department of Mathematics and Information Sciences, 
Graduate School of Science, Osaka Prefecture University, 
Nakamozu Campus, Sakai, Osaka 599-8531, Japan}
\ead{aizawa@mi.s.osakafu-u.ac.jp}

\begin{abstract}
$ {\cal N} = 2 $ Supersymmetric extensions of Galilean conformal algebra (GCA), specified by spin $ \ell $ and 
dimension of space $d$,  are investigated. 
Duval and Horv\'athy showed that the $ \ell = 1/2 $ GCA has two types of supersymmetric extensions, called 
\textit{standard} and \textit{exotic}. Recently, Masterov intorduced a centerless super-GCA for arbitrary $ \ell $ 
wchich corresponds to the standard extension. We show that the Masterov's super-GCA has  
two types of central extensions depending on the parity of $ 2\ell. $ 
We then introduced a novel super-GCA for arbitrary $ \ell $ corresponding to the exotic extension. 
It is  shown that the exotic superalgebra also has two types of  central extensions depending on the parity of $ 2\ell. $ 
Furthermore, we give a realization of the standard and exotic super-GCA's in terms of their subalgebras. 
Finally, we present a $ {\cal N} = 1 $ supersmmetric extension of GCA with central extensions. 
%
\end{abstract}

\maketitle

%
%
%
%
\section{Introduction}

 Galilean conformal algebra (GCA) is a class of non-semisimple Lie algebras. 
Its physical relevance is attributed to the fact that 
it may be ragarded as a nonrelativistic counterpart of conformal algebra \cite{NdelOR,NdelOR2}. 
Each GCA is specified by two parameters $ d \in {\mathbb N} $ and $ \ell \in \hf{\mathbb N} $ 
where $ {\mathbb N} $ denotes the set of positive integers. 
The parameter $d$ is the dimension of space in which dynamics is considered. 
This parameter $d$ also  specifies the maximal semisimple subalgebra $ sl(2,{\mathbb R}) \oplus so(d) $  of GCA.   
On the other hand the parameter $ \ell $ specifies the $ \ell d $ dimensional abelian ideal of GCA which carries the spin $ \ell $ 
representation of $ sl(2,{\mathbb R}) $ subalgebra. 
The abelian ideal has central extension for certain values of $(d,\ell) $ \cite{Henkel,StZak,LSZ,LSZ2,MT}. 

  The GCA with $ \ell = 1/2, $ the so-called Schr\"odinger algebra, is the best studied example of this class of 
algebras. One may find the Schr\"odinger algebra appearing in various fields of physics ranging from classical mechanics to 
high energy physics. 
The another member with $ \ell = 1 $ is also studied in many physical and mathematical context. 
Especially, the non-relativistic version of AdS/CFT correspondence formulated in the context of $ \ell = 1/2 $ and $ \ell = 1 $ GCA's 
\cite{MT,Son,BaMcG,BG} attracts a renewed interest on GCA's.  
Readers may refer the references in \cite{HMS,AIK} for other works on $ \ell = 1/2 $ and  $ \ell = 1 $ GCA's.  
For higher values of $ \ell, $ 
dynamical systems described by such GCA  are studied very recently in \cite{GomKam2012,AGM,GaMa,AndGo}. 
We thus think that GCA for any $ \ell $ is of physical importance.

  It is natural to consider supersymmetric extensions of the algebraic structure of 
physical importance. Several works on supersymmetrization of GCA have already been done. 
Supersymmetric extension of $ \ell = 1/2$ GCA has been completed by Duval and Horv\'athy \cite{DH}. 
In \cite{DH} a systematic method to construct  $ {\cal N}$-supersymmetric extension of 
Schr\"odinger algebras in $d$ dimensional space with central extension is presented. 
Previous to \cite{DH} there were some works on $ \ell = 1/2$ super-GCA \cite{BH,BDH,GGT,LLM}. 
These works are followed by recent active study on physical and mathematical aspects of 
$ \ell = 1/2 $ superalgebra 
\cite{NSY,HU,SY1,SY2,Nakayama,GaMa2009,Gala2009,NRSY,DoGau,BoKuPi,JKLCY,NASch1,NASch2,GoGoKa,SY3}.

For GCA with $ \ell \geq 1 $ systematic method to obtain  ${\cal N}$-supersymmetric extension  has not been established yet. 
However, there are some attempts to extend $ \ell \geq 1 $ GCA to superalgebra. 
Such an attempt was initiated by de Azc\'arraga and Lukierski \cite{AzLu}. They applyed In\"on\"u-Wigner contraction to 
$ su(2,2;{\cal N}) $ and obtained $ \ell=1 $ and $ {\cal N} = 2k $ (even) super-GCA in $(3+1)$ dimensional spacetime.      
The In\"on\"u-Wigner contraction was also used by Sakaguchi to obtain $ \ell=1 $ super-GCA in $(3+1)$ and $(2+1)$ dimensional 
spacetime \cite{Sakaguchi}. He obtained $ {\cal N} =2, 4 $ super-GCA's in $(3+1)$ dimensional 
spacetime and $ {\cal N} = 2, 4, 8 $ ones in $(2+1)$ dimensional spacetime.  
Bagchi and Mandal obtained $ \ell = {\cal N} = 1 $ super-GCA in $(3+1)$ dimensional spacetime by applying superspace contraction to 
${\cal N}=1$ extended $ so(4,2) $ \cite{BaMa}.  
A possible general structure of $ \ell = 1 $ super-GCA was discussed by Fedoruk and Lukierski \cite{FeLu}. 
Very recently, Masterov introduced $ {\cal N} = 2 $ extension of GCA for any $ d $ and $ \ell $ \cite{Masterov}. 
We remark that central extensions of $ \ell \geq 1 $ super-GCA are not considered in any works mentioned above.  

 In the present work we investigate $ {\cal N} = 2 $ supersymmetric extensions of GCA with central extension. 
The aim of this work is twofold: The first aim is to show that in the case of $ {\cal N} = 2$ 
all the supersymmetric extensions known for $ \ell = 1/2 $ is extended to arbitrary $ \ell. $ 
It is shown in \cite{DH} that $ \ell = 1/2$ GCA has two distict types of supersymmetric extensions. 
They are called \textit{standard} and \textit{exotic}. 
Masterov's $ {\cal N} = 2 $ super-GCA corresponds to the standard supersymmetric extension without central terms. 
The exotic $ {\cal N} = 2 $ super-GCA is not known yet. We thus start with Masterov's superalgebra and 
consider its central extension. Then we introduce a new superalgebra corresponding to the exotic extension 
and consider its central extension. 
In this way, we shall complete the generalization of $ \ell = 1/2 $ result to arbitrary $ \ell. $ 
Our second aim is to give a simple realization of the $ {\cal N} = 2 $ standard and exotic super-GCA's. 
It will be shonw that each super-GCA is realized on its subalgebras which is isomorphic to a boson-fermion algebra.

 This paper is organized as follows. In the next section we give a brief summary of bosnic GCA and its central 
extensions. It is pointed out that, combining the central extensions and the two types of supersymmetric extensions,  
there are four possible super-GCA's with central exension. 
In \S \ref{Sec:Mas} we sudy central extensions of the standard super-GCA by Masterov. 
We introduce the exotic super-GCA with central extension in \S \ref{Sec:Exo}. 
In \S \ref{Sec:real} we give a realization of the superalgebras introduced in \S \ref{Sec:Mas} and \S \ref{Sec:Exo}. 
This realization shows an interesting relation of the standard super-GCA and a certain bose-fermi oscillator Hamiltonian. 
Finally we introduced $ {\cal N} = 1 $ supersymmetric extension of centrally extended GCA in \S \ref{Sec:N1}. 
An another possibility of central extension is discussed in Appendix.

%
%
%
%
\section{Preliminary: Bosonic Galilean conformal algebra}
\label{Sec:Pre}

  Galilean conformal algebra for a given pair of $(d,\ell)$ has generators
\[
  D,\ H, \ C,\  M_{ij},\  \Pop{n}{i},
\]
where $ i, j = 1, 2, \cdots, d $ and $ n = 0, 1, \cdots, 2\ell. $ 
The sets of generators $ \langle D, H, C \rangle $ and $ \langle M_{ij} \rangle $ span $ sl(2,{\mathbb R}) $ and  $ so(d)  $ 
subalgebras, respectively. 
$ \Pop{n}{i} $ is a basis of abelian ideal of the GCA. The nonvanishing commutation relations 
are given by \cite{NdelOR,NdelOR2}
\begin{eqnarray}
  & & [D, H] = 2H, \qquad [D, C]=-2C, \qquad [C, H]=D,  
  \nonumber \\
  & & [M_{ij}, M_{k\ell} ] = - \delta_{ik} M_{j\ell} - \delta_{j\ell} M_{ik} 
     + \delta_{i\ell} M_{jk} + \delta_{jk} M_{i\ell},
  \nonumber \\
  & & [H, \Pop{n}{i}] = -n \Pop{n-1}{i}, \qquad\qquad 
      [D, \Pop{n}{i}] = 2(\ell-n) \Pop{n}{i}, 
  \label{BosonicDef} \\
  & & [C, \Pop{n}{i}] = (2\ell-n) \Pop{n+1}{i}, \qquad
      [M_{ij}, \Pop{n}{k} ] = -\delta_{ik} \Pop{n}{j} + \delta_{jk} \Pop{n}{i}.
  \nonumber
\end{eqnarray}
It is known that the algebra has the two types of central extensions \cite{MT}: \\
(i) mass extension existing for any $ d $ and half-integer $ \ell $
\begin{equation}
 [\Pop{m}{i}, \Pop{n}{j}] = \delta_{ij} \, \delta_{m+n,2\ell}\,  I_m M, \qquad
 I_m = (-1)^{m+\ell+\hf} (2\ell-m)! \, m!.
 \label{MassExtension}
\end{equation}
(ii) exotic extension existing only for $ d = 2 $ and integer $ \ell $
\begin{equation}
  [\Pop{m}{i}, \Pop{n}{j}] = \epsilon_{ij} \, \delta_{m+n,2\ell}\,  \check{I}_m \Theta, \qquad 
  \check{I}_m = (-1)^{m} (2\ell-m)!\,  m!, 
  \label{ExoticExtension}
\end{equation}
where $ \epsilon_{ij} $ is the antisymmetric tensor with $ \epsilon_{12} = 1. $ 
We note that this agrees, up to an overall factor, with the structure constants used in 
\cite{GomKam2012} and \cite{GalMas2011}.

  Next we briefly summarize the supersymmetric extensions of $ \ell = 1/2 $  GCA introduced by Duval and Horv\'athy.  
Supersymmetric extension requires the introduction of 
supercharges, superconformal charges and fermionic partners of $ \Pop{n}{i}. $ 
It is shown in \cite{DH} that there exist two possible ways. 
One is called \textit{standard} extension existing for any $ d, $ other is \textit{exotic} extension 
existing only for $ d = 2. $ 
One of the main differences between standard and exotic extensions is the number of 
fermionic partners of $ \Pop{n}{i}. $ If the standard extension has $ 2m$ fermionic partners, then 
the exotic extension does $ m. $ 
Combination of the central and supersymmetric extensions yields the following four possibilities of 
super-GCA with central extension:
\begin{itemize}
  \item standard super and mass central extension for any $d$ and half-integer $\ell $
  \item standard super and exotic central extension for $ d = 2 $ and integer $ \ell $
  \item exotic super and mass central extension for $ d = 2 $ and half-integer $ \ell $
  \item exotic super and exotic central extension for $ d = 2 $ and integer $ \ell $
\end{itemize}
We remark that this is not an all possible $ {\cal N} = 2 $ supersymmetric and central extensions, but 
a generalization of the result in \cite{DH} to higher $ \ell. $ 
In the following sections, we show that all four extensions are possible for $ {\cal N} = 2.$ 
This will be done based on the definition of Lie superalgebra. 
Namely, we set up a set of (anti)commutation 
relations, then verify that they satisfy the super Jacobi identity.

 Although we do not aim to full classification of all possible super and central extensions, 
one may ask the following question. The mass and exotic central extensions make the abelian ideal non-abelian. 
It is known that another type of central extension is possible for a \textit{infinite} dimensional $ \ell = 1 $ super-GCA 
in one space dimension \cite{Mandal}. Is such an central extension  possible for the \textit{finite} dimensional super-GCA ?    
This problem is studied in Appendix and we have a negative answer to this question.

%
%
%
%
\section{Standard supersymmetric extension}
\label{Sec:Mas}

$ {\cal N}=2 $ Standard supersymmetric extension of GCA for any pair of $ (d, \ell) $ 
was introduced by Masterov \cite{Masterov}. 
With a slight change of notations and conventions, the superalgebra is defined as follows. 
In addition to the generators of bosonic GCA we introduce fermionic generators 
$ Q, \bar{Q}, S, \bar{S}, \Xop{n}{i}, \Xbop{n}{i}, $ $ n = 0, 1, \cdots, 2\ell-1 $ 
and additional bosonic generators 
$ R, \Jop{n}{i}, $ $ n = 0, 1, \cdots, 2\ell-2. $ 
The suffix $i = 1, 2, \cdots, d $ denotes the space coordinate. 
Note that the generators $ \Jop{n}{i} $ does not exist for $ \ell = 1/2 $ super-GCA (super Schr\"odinger algebra). 
The generators are subjected to the nonvanishing commutation relation given in (\ref{BosonicDef}) 
and follows: The fermionic sector is given by
\begin{equation}
  \begin{array}{lcl}
     \{ Q, \bar{Q} \} = 2H, \quad \{S, \bar{S} \} = 2C, & & 
     \{ Q, \bar{S} \} = D + R, \quad \{ \bar{Q}, S \} = D -R, \\[3pt]
     \{ Q, \Xbop{n}{i} \} = -\Pop{n}{i} - n \Jop{n-1}{i}, & & 
     \{ \bar{Q}, \Xop{n}{i} \} = -\Pop{n}{i} + n \Jop{n-1}{i}, \\[3pt]
     \multicolumn{3}{l}{ \{ S, \Xbop{n}{i} \} = -\Pop{n+1}{i} - (n-2\ell+1) \Jop{n}{i},} \\[3pt]
     \multicolumn{3}{l}{ \{ \bar{S}, \Xop{n}{i} \} = -\Pop{n+1}{i} + (n-2\ell+1) \Jop{n}{i},}
  \end{array}
  \label{McentralSsuperFF}
\end{equation}
and bosonic-fermionic sector reads
\begin{equation}
  \begin{array}{llll}
    [H, S] = -Q, & [H, \bar{S}] = -\bar{Q}, & [C, Q] = S, & [C, \bar{Q}] = \bar{S}, \\[3pt]
    [D, S] = -S, & [D, \bar{S}] = -\bar{S}, & [D, Q] = Q, & [D, \bar{Q}] = \bar{Q}, \\[3pt]
    \multicolumn{2}{l}{ [H, \Xop{n}{i}] = -n\Xop{n-1}{i}, } & 
    \multicolumn{2}{l}{ [H, \Xbop{n}{i}] = -n \Xbop{n-1}{i},} \\[3pt]
    \multicolumn{2}{l}{ [D, \Xop{n}{i}] = (2\ell-2n-1) \Xop{n}{i}, } & 
    \multicolumn{2}{l}{ [D, \Xbop{n}{i}] = (2\ell-2n-1) \Xbop{n}{i},} \\[3pt]
    \multicolumn{2}{l}{ [C, \Xop{n}{i}] = (2\ell-n-1) \Xop{n+1}{i},} & 
    \multicolumn{2}{l}{ [C, \Xbop{n}{i}] = (2\ell-n-1) \Xbop{n+1}{i},} \\[3pt]
    \multicolumn{2}{l}{ [M_{ij}, \Xop{n}{k}] = -\delta_{ik}\Xop{n}{j} + \delta_{jk} \Xop{n}{i},} & 
    \multicolumn{2}{l}{ [M_{ij}, \Xbop{n}{k}] = -\delta_{ik} \Xbop{n}{j} + \delta_{jk} \Xbop{n}{i},} \\[3pt]
    [R,Q] = -Q, & [R, \bar{Q}]= \bar{Q}, & [R, S] = -S, & [R, \bar{S}]=\bar{S}, \\[3pt]
    [R, \Xop{n}{i}]=-\Xop{n}{i},& & [R, \Xbop{n}{i}] = \Xbop{n}{i}, \\[3pt]
    \multicolumn{2}{l}{ [Q, \Pop{n}{i}] = n \Xop{n-1}{i},} & 
    \multicolumn{2}{l}{ [\bar{Q}, \Pop{n}{i}] = n \Xbop{n-1}{i},} \\[3pt]
    \multicolumn{2}{l}{  [Q, \Jop{n}{i}]=-\Xop{n}{i},} & \multicolumn{2}{l}{ [\bar{Q}, \Jop{n}{i}] = \Xbop{n}{i},}
    \\[3pt]
    \multicolumn{2}{l}{ [S, \Pop{n}{i} ]= -(2\ell-n) \Xop{n}{i},} & 
    \multicolumn{2}{l}{ [\bar{S}, \Pop{n}{i}] = -(2\ell-n) \Xbop{n}{i},} \\[3pt]
    \multicolumn{2}{l}{ [S, \Jop{n}{i}]= -\Xop{n+1}{i},} & 
    \multicolumn{2}{l}{ [\bar{S}, \Jop{n}{i}] = \Xbop{n+1}{i}.}
  \end{array}
  \label{McentralSsuperBF}
\end{equation}
Finally the additional bosonic generators satisfy the relations
\begin{equation}
  \begin{array}{ll}
     [H, \Jop{n}{i}] = -n \Jop{n-1}{i}, & [D, \Jop{n}{i}]= 2(\ell-n-1) \Jop{n}{i}, \\[3pt]
     [C, \Jop{n}{i}] = (2\ell-n-2) \Jop{n+1}{i}, & [M_{ij}, \Jop{n}{k}] = -\delta_{ik} \Jop{n}{j} + \delta_{jk} \Jop{n}{i}.
  \end{array}
  \label{McentralSsuperBB}
\end{equation}
As seen from these commutation relations, $ \langle Q, \bar{Q}, S, \bar{S}, D, H, C, R \rangle $ spans 
$ sl(2/1) $ subalgebra and $ \langle \Pop{n}{i}, \Xop{n}{i}, \Xbop{n}{i}, \Jop{n}{i} \rangle $ forms an abelian subalgebra.

 For half-integral values of $ \ell $ one may verify that this standard super-GCA 
has the mass central extension defined by (\ref{MassExtension}) and 
\begin{equation}
 \{ \Xop{m}{i}, \Xbop{n}{j} \} = \delta_{ij} \delta_{m+n, 2\ell-1} \alpha_m M, \quad
 [\Jop{m}{i}, \Jop{n}{j}] = \delta_{ij} \delta_{n+m,2\ell-2} \beta_m M,
  \label{MassCE}
\end{equation}
where
\begin{equation}
  \alpha_m = (-1)^{m+\ell-\hf} (2\ell-1-m)! \, m!, \quad 
  \beta_m  = (-1)^{m+\ell+\hf} (2\ell-2-m)! \, m!.
  \label{Ialpha}
\end{equation}
For $ d = 2 $ and integral values of  $ \ell $ one may also verify that it has the exotic central extension. 
Since the dimension of space is two, the rotation subalgebra is $ so(2) $ generated by 
only one element $ M_{12}. $ The additional nonvanishing commutators are given by  
(\ref{ExoticExtension}) and 
\begin{equation}
 \{ \Xop{m}{i}, \Xbop{n}{j} \} = \epsilon_{ij} \delta_{m+n, 2\ell-1} \check{\alpha}_m \Theta,
 \quad
 [\Jop{m}{i}, \Jop{n}{j} ]= \epsilon_{ij} \delta_{m+n,2\ell-2} \check{\beta}_m \Theta,
 \label{SsuperEcentral}
\end{equation}
where
\begin{equation}
   \check{\alpha}_m = (-1)^{m+1} (2\ell-1-m)! \, m!, \quad
   \check{\beta}_m = (-1)^m m!\, (2\ell-2-m)!. 
   \label{checkab}
\end{equation}

%
%
%
%
\section{Exotic supersymmetric extension}
\label{Sec:Exo}

  In this section, we define $ {\cal N} = 2 $  exotic supersymmetric extension of GCA. 
As such extension for $ \ell = 1/2 $ exists only in two dimensional space, we set  $ d = 2. $ 
The space rotation is generated by one element $ M_{12}. $ 
We introduce fermionic generators $ Q, Q^*, S, S^*, \Xop{n}{i}, $ $ n = 0, 1, \cdots, 2\ell-1 $ 
and an additional bosonic generator $ R. $ 
In comparison with the standard super-GCA in the previous section, 
the number of the $ X$-type generator is half and there is no $\Jop{n}{i}. $ 
We set the following nonvanishing commutators as well as (\ref{BosonicDef}). 
the fermionic sector:
\begin{equation}
  \begin{array}{lcl}
    \{ Q, Q \} = \{ Q^*, Q^* \} = 2 H, & & \{ S, S \} = \{ S^*, S^* \} = 2C, \\[3pt]
    \{ Q, S \} = \{ Q^*, S^* \} = D, \\[3pt]
    \{ Q, S^* \} = -M_{12} + R, & &  \{ Q^*, S \} = M_{12} - R, \\[3pt] 
    \{ Q, \Xop{n}{i} \} = -\Pop{n}{i},  & &  \displaystyle \{ Q^*, \Xop{n}{i} \} = \sum_k \epsilon_{ik} \Pop{n}{k}, \\[3pt]
    \{ S, \Xop{n}{i} \} = -\Pop{n+1}{i}, & & \displaystyle \{ S^*, \Xop{n}{i} \} = \sum_k \epsilon_{ik} \Pop{n+1}{k}, 
  \end{array}
  \label{McentralEsuperFF}
\end{equation}
bosonic-fermionic sector:
\begin{equation}
  \begin{array}{llll}
    [H, S] = -Q, & [H, S^*] = -Q^*, &  [C, Q] = S, & [C, Q^*] = S^*, \\[3pt]
    [D, S] = -S, & [D, S^*] = -S^*, &  [D, Q] = Q, & [D, Q^*] = Q^*, \\[3pt]
    \multicolumn{2}{l}{ [H, \Xop{n}{i}] = -n \Xop{n-1}{i},} & \multicolumn{2}{l}{ [C, \Xop{n}{i}] = (2\ell-n-1) \Xop{n+1}{i},} \\[3pt]
    \multicolumn{2}{l}{ [D, \Xop{n}{i}] = (2 \ell - 2n -1) \Xop{n}{i},}  & 
    \multicolumn{2}{l}{\displaystyle [R, \Xop{n}{i}] = (2\ell+1) \sum_k \epsilon_{ik} \Xop{n}{k},} \\[3pt]
    [M_{12}, S] = S^*, & [M_{12}, S^*] = -S,  & [M_{12}, Q] = Q^*, & [M_{12}, Q^*] = -Q, \\[3pt]
    [R, S] = 2 S^*, & [R, S^*] = -2S,  & [R, Q] = 2Q^*, & [R, Q^*]=-2Q, \\[3pt]
    \multicolumn{2}{l}{ [Q, \Pop{n}{i}] = n \Xop{n-1}{i}, } & 
    \multicolumn{2}{l}{ \displaystyle [Q^*, \Pop{n}{i} ] = n \sum_k \epsilon_{ik} \Xop{n-1}{k}, } \\[3pt]
    \multicolumn{2}{l}{ [S, \Pop{n}{i}] = -(2\ell-n) \Xop{n}{i}, } & 
    \multicolumn{2}{l}{ \displaystyle [S^*, \Pop{n}{i}] = -(2\ell-n) \sum_k \epsilon_{ik} \Xop{n}{k}.}
  \end{array}
  \label{McentralEsuperBF}
\end{equation}
additional bosonic commutator:
\begin{equation}
  [R, \Pop{n}{i}] = (2\ell-1) \sum_k \epsilon_{ik} \Pop{n}{k}.
  \label{McentralEsuperBB}
\end{equation}
These relations define the $ {\cal N} = 2 $ exotic super-GCA without central extension as they satisfy the super 
Jacobi identity. This super-GCA has $ osp(2/2) $ subalgebra generated by $ \langle Q, Q^*, S, S^*, D, H, C, R \rangle. $ 
The set of generators $ \langle \Pop{n}{i}, \Xop{n}{i} \rangle $ forms an abelian ideal. 

  Now we consider central extensions of the exotic super-GCA. 
  For half-integer $ \ell $ we introduce the mass central extension defined by (\ref{MassExtension}) 
and
\begin{equation}
 \{ \Xop{m}{i}, \Xop{n}{j} \} = \delta_{ij} \delta_{m+n, 2\ell-1} \alpha_m M,  \quad
 \alpha_m = (-1)^{m+\ell-\hf} (2\ell-1-m)! \, m!.
  \label{MassCE0}
\end{equation}
For integer $ \ell $ we introduce the exotic central extension by (\ref{ExoticExtension}) and 
\begin{equation}
 \{ \Xop{m}{i}, \Xop{n}{j} \} = \epsilon_{ij} \delta_{m+n, 2\ell-1} \check{\alpha}_m \Theta, 
 \quad
 \check{\alpha}_m = (-1)^{m+1} (2\ell-1-m)! \, m!.
  \label{ExoticCE}
\end{equation}
One may verify by the straightforward computation 
that these commutation relations satisfy the super Jacobi identity. 
Thus we have obtained the $ {\cal N} = 2 $ exotic super-GCA with mass or exotic central extensions. 

%
%
%
%
\section{Realizations}
\label{Sec:real}

 In \cite{AGM} dynamical systems on which bosonic GCA with mass or exotic central extension 
acts as symmetry operation were constructed by the orbit method. The construction shows as a 
byproduct that the bosonic GCA can be realized in terms of $ \Pop{n}{i}. $ 
As seen in \S \ref{Sec:Pre}  the central extensions make $ \Pop{n}{i} $ noncommutative so that we may use 
them to realize the GCA. 
In this section we generalize this realization to the super-GCA's introduced in the previous sections. 
The obtained is a realization of the abstract Lie superalgebra by its subalgebra. 
However, this also leads an interesting relation of the super-GCA and a bose-fermi  Hamiltonian.  
In the realization we replace the central elements $M, \Theta $ with their 
eigenvalues $ \mu, \theta, $ respectively. We use the vector notation such as 
$ \bi{P}^{(n)} = ( \Pop{n}{1}, \Pop{n}{2}, \cdots, \Pop{n}{d}). $

\subsection{standard super and mass central extension}

The super-GCA defined by (\ref{BosonicDef}), (\ref{MassExtension}), (\ref{McentralSsuperFF}), (\ref{McentralSsuperBF}), 
(\ref{McentralSsuperBB}) and (\ref{MassCE}) is realized in terms of  $ \Pop{n}{i}, \Xop{n}{i}, \Xbop{n}{i}, \Jop{n}{i}. $ 
It is given by the following equations:
\begin{eqnarray}
  & & H = \frac{1}{2\mu} \left(
          \sum_{m=1}^{2\ell} \frac{m}{I_m} \bi{P}^{(2\ell-m)} \bi{P}^{(m-1)} 
        + 2 \sum_{m=1}^{2\ell-1} \frac{m}{\alpha_m} \bi{X}^{(2\ell-1-m)} \bar{\bi{X}}^{(m-1)} 
        \right.
  \nonumber \\
  & & \qquad \qquad \left.
       + \sum_{m=1}^{2\ell-2} \frac{m}{\beta_m}  \bi{J}^{(2\ell-2-m)} \bi{J}^{(m-1)})
       \right),
  \nonumber \\
  & & D = \frac{1}{\mu} \left(
        \sum_{m=0}^{2\ell} \frac{ m-\ell }{ I_m }  \bi{P}^{(2\ell-m)} \bi{P}^{(m)}
        + \sum_{m=0}^{2\ell-1} \frac{2m+1-2\ell }{ \alpha_m } \bi{X}^{(2\ell-1-m)} \bar{\bi{X}}^{(m)} 
        \right.
  \nonumber \\
  & & \qquad \qquad \left.
        + \sum_{m=0}^{2\ell-2} \frac{m+1-\ell }{ \beta_m } \bi{J}^{(2\ell-2-m)} \bi{J}^{(m)}
       \right),
  \nonumber \\
  & & C = \frac{1}{2\mu} \left( 
          \sum_{m=1}^{2\ell} \frac{m}{I_m} \bi{P}^{(2\ell+1-m)} \bi{P}^{(m)} 
        + 2 \sum_{m=1}^{2\ell-1} \frac{m}{\alpha_m} \bi{X}^{(2\ell-m)} \bar{\bi{X}}^{(m)}          
        \right.
  \nonumber \\
  & & \qquad \qquad \left.
       + \sum_{m=1}^{2\ell-2} \frac{m}{\beta_m}  \bi{J}^{(2\ell-1-m)} \bi{J}^{(m)}
       \right),
  \nonumber \\
  & & M_{ij} = \frac{1}{2\mu} \left( 
         \sum_{m=0}^{2\ell} \frac{1}{ I_m } ( \Pop{2\ell-m}{i} \Pop{m}{j} - \Pop{2\ell-m}{j} \Pop{m}{i} )
        \right.
  \nonumber \\
  & & \qquad \qquad \left.
       + 2 \sum_{m=0}^{2\ell-1} \frac{1}{\alpha_m} ( \Xop{2\ell-1-m}{i} \Xbop{m}{j} + \Xbop{m}{i} \Xop{2\ell-1m}{j} )
      \right.
  \nonumber \\
  & & \qquad \qquad \left.
       + \sum_{m=0}^{2\ell-2} \frac{1}{\beta_m} ( \Jop{2\ell-2-m}{i} \Jop{m}{j} - \Jop{2\ell-2-m}{j} \Jop{m}{i} )
       \right),
  \label{RelMcenterSsuper} \\
  & & R = \frac{-1}{\mu} \sum_{m=0}^{2\ell-1} \frac{1}{\alpha_m} \bi{X}^{(2\ell-1-m)} \bar{\bi{X}}^{(m)} + \ell d,
  \nonumber \\
  & & Q = \frac{-1}{\mu} \sum_{m=1}^{2\ell} \frac{m}{I_m} ( \bi{P}^{(2\ell-m)} + (2\ell-m) \bi{J}^{(2\ell-1-m)} ) \bi{X}^{(m-1)},
  \nonumber \\
  & & \bar{Q} = \frac{-1}{\mu} \sum_{m=1}^{2\ell} \frac{m}{I_m} ( \bi{P}^{(2\ell-m)} - (2\ell-m) \bi{J}^{(2\ell-1-m)} ) \bar{\bi{X}}^{(m-1)},
  \nonumber \\
  & & S = \frac{-1}{\mu} \sum_{m=1}^{2\ell} \frac{m}{I_m} ( \bi{P}^{(2\ell+1-m)} - (m-1) \bi{J}^{(2\ell-m)} ) \bi{X}^{(m-1)},
  \nonumber \\
  & & \bar{S} = \frac{-1}{\mu} \sum_{m=1}^{2\ell} \frac{m}{I_m} ( \bi{P}^{(2\ell+1-m)} + (m-1) \bi{J}^{(2\ell-m)} ) \bar{\bi{X}}^{(m-1)}.
  \nonumber 
\end{eqnarray}
The mass central extension makes the subalgebra $\langle  \Pop{n}{i}, \Xop{n}{i},  \Jop{n}{i} \rangle $ noncommutative. 
This noncommutative subalgebra is isomorphic to the boson-fermion algebra. 
Therefore the realization (\ref{RelMcenterSsuper}) is nothing but a boson-fermion realization of super-GCA. 
Let us introduce bosons and fermions by which we may rewrite (\ref{RelMcenterSsuper}).  
We define $ 2\ell $ kinds of bosons in $(d+1)$ dimensional spacetime: 
\begin{equation}
  b_i^{(n)} = \left\{
    \begin{array}{lcl}
        \frac{ 1}{ \sqrt{\mu} I_{n-1} } \Pop{n-1}{i} , & & n = 1, 2, \cdots, \ell + \hf \\[7pt]
        \frac{ 1}{ \sqrt{\mu} \beta_{n-\ell- \frac{3}{2}} } \Jop{n-\ell- \frac{3}{2}}{i} , & & 
           n = \ell+\frac{3}{2}, \cdots, 2\ell
    \end{array}
  \right.
  \label{Boson-ann}
\end{equation}
\begin{equation}
  b_i^{\dagger(n)} = \left\{
    \begin{array}{lcl}
        \frac{ 1}{ \sqrt{\mu}  } \Pop{2\ell-n+1}{i} , & & n = 1, 2, \cdots, \ell + \hf \\[7pt]
        \frac{ 1}{ \sqrt{\mu}  } \Jop{3\ell-n-\hf}{i} , & & 
           n = \ell+\frac{3}{2}, \cdots, 2\ell
    \end{array}
  \right.
  \label{Boson-cre}
\end{equation}
We also introduce $ 2\ell $ kinds of fermion:   
\begin{equation}
    \alpha_i^{(n)} = \frac{-1}{ \sqrt{\mu} \alpha_{n-1} } \Xbop{n-1}{i}, \qquad 
    \alpha_i^{\dagger(n)} = \frac{1}{\sqrt{\mu}} \Xop{2\ell-n}{i},
    \label{Fermion-def}
\end{equation}
where $ n = 1, 2, \cdots, 2\ell. $ Then they satisfy the commutation relations:
\begin{equation}
  [b^{(m)}_i, b^{\dagger(n)}_j] = \delta_{ij} \, \delta_{mn}, \qquad 
  \{ \alpha^{(m)}_i, \alpha^{\dagger(n)}_j \} = \delta_{ij} \, \delta_{mn}.
  \label{boson-fermion-comm}
\end{equation}
It is obvious that one can rewrite the realization (\ref{RelMcenterSsuper}) with these bosons and fermions. 
One may recognize that the generator $ D, $ in the realization by the bosons and fermions, is a linear combination 
of bose and femi oscillators. Then it relates to the 
following bose-fermi oscillator Hamiltonian
\begin{eqnarray}
  {\cal H} &=& \sum_{m=1}^{\ell+\hf} (\ell-m+1) \bi{b}^{\dagger(m)} \bi{b}^{(m)}
           + \sum_{m=\ell+\frac{3}{2}}^{2\ell} \left(2\ell-m+\hf \right) \bi{b}^{\dagger(m)} \bi{b}^{(m)}
     \nonumber \\
           &-& \sum_{m=1}^{2\ell} \left(\ell-m+\hf \right) \balpha^{\dagger(m)} \balpha^{(m)}.
      \label{BFHamiltonian}
\end{eqnarray}
The relations is given by
\[
 D = -2 {\cal H} - \ell \left( \ell + \hf \right) d,
\]
namely, one may interpret $ D $ as a Hamiltonian. 
Then each generator of the standard super-GCA is either a raising operator or a lowering operator or 
a symmetry operator of the eigenvalue of $ {\cal H}. $ 
This interpretation is different from the ordinary supersymmetric theories where 
the generator $H$ plays the role of Hamiltonian. 
However, we think that this is an interesting observation resulting from the realization (\ref{RelMcenterSsuper}).

\subsection{standard super and exotic central extension}

The super-GCA defined by (\ref{BosonicDef}), (\ref{ExoticExtension}), (\ref{McentralSsuperFF}), (\ref{McentralSsuperBF}), 
(\ref{McentralSsuperBB}) and (\ref{SsuperEcentral}) is realized in terms of  $ \Pop{n}{i}, \Xop{n}{i}, \Xbop{n}{i}, \Jop{n}{i}. $ 
It is given by the following equations:
\begin{eqnarray}
  & & H = \frac{-1}{2\theta} \sum_{j,k=1}^2 \epsilon_{jk} \left(
        \sum_{m=1}^{2\ell} \frac{m}{\check{I}_m}  \Pop{2\ell-m}{j} \Pop{m-1}{k}
        + 2 \sum_{m=1}^{2\ell-1} \frac{m}{\check{\alpha}_m}  \Xop{2\ell-1-m}{j} \Xbop{m-1}{k}
      \right.
      \nonumber \\
  & & \qquad \qquad \left.
       + \sum_{m=1}^{2\ell-2} \frac{m}{\check{\beta}_m}  \Jop{2\ell-2-m}{j} \Jop{m-1}{k}
      \right),
      \nonumber \\
  & & D = \frac{1}{\theta} \sum_{j,k=1}^2 \epsilon_{jk} \left(
      \sum_{m=0}^{2\ell} \frac{\ell- m}{\check{I}_m}  \Pop{2\ell-m}{j} \Pop{m}{k}
      + \sum_{m=0}^{2\ell-1} \frac{2\ell-2m-1}{\check{\alpha}_m}  \Xop{2\ell-1-m}{j} \Xbop{m}{k}
      \right.
      \nonumber \\
  & & \qquad \qquad \left.
      + \sum_{m=1}^{2\ell-2} \frac{\ell-m-1}{\check{\beta}_m}  \Jop{2\ell-2-m}{j}\Jop{m}{k}
      \right),
      \nonumber \\
  & & C = \frac{-1}{2\theta} \sum_{j,k=1}^2 \epsilon_{jk} \left(
      \sum_{m=1}^{2\ell} \frac{m}{\check{I}_m}  \Pop{2\ell+1-m}{j} \Pop{m}{k} 
      + 2 \sum_{m=1}^{2\ell-1} \frac{m}{\check{\alpha}_m}  \Xop{2\ell-m}{j} \Xbop{m}{k}
      \right.
      \nonumber \\
  & & \qquad \qquad \left.
      + \sum_{m=1}^{2\ell-2} \frac{m}{\check{\beta}_m} \Jop{2\ell-1-m}{j} \Jop{m}{k}
      \right),
      \nonumber \\
  & & M_{12} = \frac{1}{2\theta} \left(
      \sum_{m=0}^{2\ell} \frac{1}{\check{I}_m} \bi{P}^{(2\ell-m)} \bi{P}^{(m)} 
         + 2 \sum_{m=0}^{2\ell-1} \frac{1}{ \check{\alpha}_m } \bi{X}^{(2\ell-1-m)} \bar{\bi{X}}^{(m)}
      \right.
      \label{RelEcenterSsuper} \\
  & & \qquad \qquad \left.
         + \sum_{m=0}^{2\ell-2} \frac{1}{\check{\beta}_m} \bi{J}^{(2\ell-2-m)} \bi{J}^{(m)}
      \right),
      \nonumber \\
  & & R = \frac{1}{\theta} \sum_{j,k=1}^2 \sum_{m=0}^{2\ell-1} 
          \frac{1}{ \check{\alpha}_m } \epsilon_{jk} \Xop{2\ell-1-m}{j} \Xbop{m}{k} 
          + 2 \ell,
      \nonumber \\
  & & Q = \frac{1}{\theta} \sum_{j,k=1}^2 \sum_{m=1}^{2\ell} \frac{m}{ \check{I}_m } \epsilon_{jk}  
          ( \Pop{2\ell-m}{j} + (2\ell-m) \Jop{2\ell-1-m}{j} ) \Xop{m-1}{k},
      \nonumber \\
  & & \bar{Q} = \frac{1}{\theta} \sum_{j,k=1}^2 \sum_{m=1}^{2\ell} \frac{m}{ \check{I}_m } \epsilon_{jk}  
          ( \Pop{2\ell-m}{j} - (2\ell-m) \Jop{2\ell-1-m}{j} ) \Xbop{m-1}{k},
      \nonumber \\
  & & S = \frac{1}{\theta} \sum_{j,k=1}^2 \sum_{m=1}^{2\ell} \frac{m}{ \check{I}_m } \epsilon_{jk}  
          ( \Pop{2\ell+1-m}{j} - (m-1) \Jop{2\ell-m}{j} ) \Xop{m-1}{k},
      \nonumber \\
  & & \bar{S} = \frac{1}{\theta} \sum_{j,k=1}^2 \sum_{m=1}^{2\ell} \frac{m}{ \check{I}_m } \epsilon_{jk}  
          ( \Pop{2\ell+1-m}{j} + (m-1) \Jop{2\ell-m}{j} ) \Xbop{m-1}{k}.
      \nonumber 
\end{eqnarray}

\subsection{exotic super and mass central extension}

The super-GCA defined by (\ref{BosonicDef}), (\ref{MassExtension}) and (\ref{McentralEsuperFF})-(\ref{MassCE0}) 
is realized in terms of  $ \Pop{n}{i}, \Xop{n}{i}. $  
It is given by the following equations:
\begin{eqnarray}
  & & H = \frac{1}{2\mu} \left(
        \sum_{m=1}^{2\ell} \frac{m}{I_m}  \bi{P}^{(2\ell-m)} \bi{P}^{(m-1)} 
        + \sum_{m=1}^{2\ell-1} \frac{m}{\alpha_m}\bi{X}^{(2\ell-1-m)} \bi{X}^{(m-1)}
      \right),
 \nonumber \\
 & & D  = \frac{1}{\mu} 
        \sum_{m=0}^{2\ell} \frac{m-\ell}{I_m}\bi{P}^{(2\ell-m)} \bi{P}^{(m)}
        + \frac{1}{2\mu} \sum_{m=0}^{2\ell-1} \frac{2m + 1 - 2\ell}{\alpha_m} \bi{X}^{(2\ell-1-m)} \bi{X}^{(m)},
 \nonumber \\
 & & C = \frac{1}{2\mu} \left(
       \sum_{m=1}^{2\ell} \frac{m}{I_m}  \bi{P}^{(2\ell+1-m)} \bi{P}^{(m)} 
        + \sum_{m=1}^{2\ell-1} \frac{m}{\alpha_m}\bi{X}^{(2\ell-m)} \bi{X}^{(m)}
     \right),
 \nonumber \\
 & & M_{12} = \frac{1}{2\mu} \sum_{m=0}^{2\ell} \sum_{j,k=1}^2  \frac{\epsilon_{jk}}{I_m} \Pop{2\ell-m}{j} \Pop{m}{k},
  \label{RelMcenterEsuper} \\
 & & R = -\frac{2 \ell + 1}{2 \mu} \sum_{m=0}^{2\ell-1} \sum_{j,k=1}^2  \frac{ \epsilon_{jk} }{\alpha_m} 
        \Xop{2\ell-1-m}{j} \Xop{m}{k} - (2\ell-1) M_{12},
 \nonumber \\
 & & Q = \frac{-1}{\mu}  \sum_{m=1}^{2\ell} \frac{m}{I_m} \bi{P}^{(2\ell-m)} \bi{X}^{(m-1)},
 \nonumber \\
 & & Q^* = \frac{-1}{\mu}  \sum_{m=1}^{2\ell} \sum_{j,k=1}^2  \frac{m}{I_m} \epsilon_{jk} \Pop{2\ell-m}{j} \Xop{m-1}{k},
 \nonumber \\
 & & S = \frac{-1}{\mu}  \sum_{m=1}^{2\ell} \frac{m}{I_m} \bi{P}^{(2\ell+1-m)} \bi{X}^{(m-1)},
 \nonumber \\
 & & S^* = \frac{-1}{\mu}  \sum_{m=1}^{2\ell} \sum_{j,k=1}^2 \frac{m}{I_m} \epsilon_{jk}  \Pop{2\ell+1-m}{j} \Xop{m-1}{k}.
 \nonumber
\end{eqnarray}

\subsection{exotic super and exotic central extension}

The super-GCA defined by (\ref{BosonicDef}), (\ref{ExoticExtension}), (\ref{McentralEsuperFF})-(\ref{McentralEsuperBB}) 
and (\ref{ExoticCE}) is realized in terms of  $ \Pop{n}{i}, \Xop{n}{i}. $  
It is given by the following equations: 
\begin{eqnarray}
  & &  H = \frac{-1}{2\theta} \sum_{j,k=1}^2 \epsilon_{jk} \left(
           \sum_{m=1}^{2\ell} \frac{m}{\check{I}_m}  \Pop{2\ell-m}{j} \Pop{m-1}{k}
         + \sum_{m=1}^{2\ell-1} \frac{m}{\check{\alpha}_m} \Xop{2\ell-1-m}{j} \Xop{m-1}{k}
       \right),
  \nonumber \\
  & & D = \sum_{j,k=1}^2 \epsilon_{jk} \left( 
         \frac{1}{\theta}  \sum_{m=0}^{2\ell} \frac{\ell-m}{\check{I}_m}  \Pop{2\ell-m}{j} \Pop{m}{k} 
        + \frac{1}{2\theta} \sum_{m=0}^{2\ell-1} \frac{2\ell-2m-1}{\check{\alpha}_m} \Xop{2\ell-1-m}{j} \Xop{m}{k}
       \right),
  \nonumber \\
  & & C = \frac{-1}{2\theta}  \sum_{j,k=1}^2 \epsilon_{jk} \left(
           \sum_{m=1}^{2\ell} \frac{m}{\check{I}_m}  \Pop{2\ell+1-m}{j} \Pop{m}{k}
         + \sum_{m=1}^{2\ell-1} \frac{m}{\check{\alpha}_m} \Xop{2\ell-m}{j} \Xop{m}{k}
       \right),
  \nonumber \\
 & & M_{12} = \frac{1}{2\theta} \sum_{m=0}^{2\ell} \frac{1}{\check{I}_m} \bi{P}^{(2\ell-m)} \bi{P}^{(m)},
 \label{RelEcenterEsuper} \\  
 & & R = -\frac{ 2\ell+1 }{ 2\theta } \sum_{m=0}^{2\ell-1}  \frac{1}{ \check{\alpha}_m } \bi{X}^{(2\ell-1-m)} \bi{X}^{(m)}
        - (2\ell-1) M_{12},
  \nonumber \\
 & & Q = \frac{1}{\theta} \sum_{m=1}^{2\ell} \sum_{j,k=1}^2 \frac{ m }{\check{I}_m}  \epsilon_{jk}  \Pop{2\ell-m}{j} \Xop{m-1}{k},
  \nonumber \\
 & & Q^* = \frac{-1}{\theta} \sum_{m=1}^{2\ell}  \frac{ m }{\check{I}_m}\bi{P}^{(2\ell-m)} \bi{X}^{(m-1)},
  \nonumber \\
 & & S = \frac{1}{\theta} \sum_{m=1}^{2\ell} \sum_{j,k=1}^2 \frac{ m }{\check{I}_m}  \epsilon_{jk}  \Pop{2\ell+1-m}{j} \Xop{m-1}{k},
  \nonumber \\
 & & S^* = \frac{-1}{\theta} \sum_{m=1}^{2\ell}  \frac{ m }{\check{I}_m}\bi{P}^{(2\ell+1-m)} \bi{X}^{(m-1)},
  \nonumber 
\end{eqnarray}

%
%
%
%
\section{$ {\cal N} = 1 $ supersymmetric extension}
\label{Sec:N1}

In this section we give a $ {\cal N} = 1 $ supersymmetric extension of GCA for any $ (d,\ell) $ 
and their central extensions. We remark that $ {\cal N} = 1 $ extension is not 
obtained by In\"on\"u-Wigner contraction \cite{AzLu,Sakaguchi}. 
$ {\cal N} = 1 $ super-GCA for $ \ell = 1 $ is discussed in \cite{BaMa,FeLu}, 
however central extension is not considered. 

 We introduce fermionic generators $ Q, S $ and $  \Xop{n}{i}, $ $ n = 0, 1, \cdots, 2\ell-1 $ 
but no additional bosonic ones.  
Then we set up the following anti-commutation relations for fermionic generators:
\begin{equation}
   \begin{array}{lclcl}
     \{ Q, Q \} = 2H, & & \{ S, S \} = 2C, & & \{ Q, S \} = D, \\[3pt]
     \multicolumn{2}{l}{ \{ Q, \Xop{n}{i} \} = -\Pop{n}{i}, } & 
     \multicolumn{2}{l}{ \{ S, \Xop{n}{i} \} = -\Pop{n+1}{i}, } & 
   \end{array}
   \label{N=1massF}
\end{equation}
and the commutation relations for pairs of bosonic and fermionic generators:
\begin{equation}
   \begin{array}{llll}
      [H, S] = -Q, & [C, Q] = S, & [D, S] = -S, & [D, Q] = Q, \\[3pt]
      \multicolumn{2}{l}{ [H, \Xop{n}{i} ] = -n \Xop{n-1}{i}, } & 
      \multicolumn{2}{l}{ [D, \Xop{n}{i} ] = (2\ell-2n-1) \Xop{n}{i}, } \\[3pt]
      \multicolumn{2}{l}{ [C, \Xop{n}{i}] = (2\ell-n-1) \Xop{n+1}{i},} & 
      \multicolumn{2}{l}{ [M_{ij}, \Xop{n}{k} ] = - \delta_{ik} \Xop{n}{j} + \delta_{jk} \Xop{n}{i},}
      \\[3pt]
      [Q, \Pop{n}{i}] = n \Xop{n-1}{i}, & & 
      \multicolumn{2}{l}{ [S, \Pop{n}{i}] = -(2\ell-n) \Xop{n}{i}.   }
   \end{array}
   \label{N=1massFB}
\end{equation}
We assume that other (anti)-commutators vanish. Then one can verify that these relations 
are compatible with super Jacobi identity so that define a $ {\cal N} = 1 $ super-GCA 
without central extensions. 

 Now we turn to central extensions. 
For half-integer $ \ell $ and any $d$ we may introduce the mass central extension 
given by (\ref{MassCE0}). While for integer $ \ell $ and $ d = 2 $ we introduce 
the exotic central extension given by  (\ref{ExoticCE}). 
It is not difficult to verify that these central extensions do not contradict with the super Jacobi 
identity. Therefore we have obtained the $ {\cal N} = 1 $ super-GCA with mass or 
exotic central extension.

%
%
%
%
\section{Concluding remarks} 

  We discussed $ {\cal N} = 2 $ supersymmetric extensions of GCA and their central extensions. 
We have shown that the supersymmetric and the central extensions of $ \ell = 1/2 $ GCA presented in \cite{DH} 
is able to generalize to higher $ \ell. $ 
We do not claim that we give a exhaustive list of $ {\cal N} = 2 $ super-GCA. 
For instance, it is shown in  \cite{DH} that for a given $ {\cal N} $ the supersymmetric extension of $ \ell = 1/2$ GCA is 
determined by a pair of positive integers $ (N_+, N_-) $ satisfying $ N_+ + N_- = {\cal N}. $ 
The $ {\cal N} = 2 $ extensions presented in this paper correspond to $ N_+ = 2, N_- = 0. $ 
Thus super-GCA correspond to other pairs of $ (N_+, N_-) $ may exist. 

 Many works on supersymmetric extension of $ \ell \geq 1 $ GCA focus on the study of algebraic structure. 
Although a relation between $ \ell =1 $ super-GCA and  superconformal mechanics is discussed in \cite{FeLu},
we know very few on physical implication or representation theory of $ \ell \geq 1 $ super-GCA. 
The bose-fermi oscillator Hamiltonian given in \S \ref{Sec:real} provides another physical model 
relating to super-GCA. However, the physical application of $ \ell \geq 1 $  super-GCA 
is still an open problem. Representation theory of  super-GCA is also a  
problem to be studied. We give a realization of the super-GCA in \S \ref{Sec:real}. 
However, more mathematical works such as classification of irreducible representations 
(see \cite{NASch1,NASch2} for $ \ell = 1/2 $) should be done for further understanding 
and physical applications of super-GCA. 

   We close this paper by mentioning that super-GCA is easily extended to 
infinite dimensional algebras \cite{HU,BaMa,Masterov,Mandal}. 
They have (super) Virasoro algebra as a subalgebra. Because of this, we think that 
infinite dimensional extension of the superalgebras introduced in this paper may be an 
interesting object from mathematical and physical point of view.

%
%
\ack{This work is supported by a grants-in-aid from JSPS (Contract No.23540154). }

\appendix
\section*{Appendix}
\setcounter{section}{1}

 The central extensions considered in \S \ref{Sec:Mas} and \S \ref{Sec:Exo} make the abelian 
ideal of super-GCA non-abelian. In this appendix we study an another possibility of central extension 
inspired from the infinite dimensional super-GCA. 
In \cite{Mandal} an infinite dimensional super-GCA corresponding to $ d = \ell = 1 $ is considered and 
the superalgebra has the central extension different from those in  \S \ref{Sec:Mas} and \S \ref{Sec:Exo}. 
We study the possibility of such a central extension for the finite dimensional super-GCA considered in 
the present work.  To keep contact with the infinite dimensional algebra in \cite{Mandal} $ d $ is set to be 1  
but $ \ell $ remains arbitrary. Thus we start with the ${\cal N} = 2 $ standard supersymmetric extension of GCA 
without central extension. 
Since it has the  subalgebra $ sl(2/1) $ generated by $ \langle Q, \bar{Q}, S, \bar{S}, D, H, C, R \rangle $ 
we introduce the following central terms:
\begin{eqnarray}
  & & [D, P^{(n)}] = 2(\ell-n) P^{(n)} + \alpha^{(n)}, \quad
      [H, P^{(n)}] = -n P^{(n-1)} + \beta^{(n)}, \nonumber
  \\
  & & [C, P^{(n)}] = (2\ell-n) P^{(n+1)} + \gamma^{(n)}, \nonumber
  \\
  & & \{ Q, \bar{X}^{(n)} \} = -P^{(n)}-n J^{(n-1)} + c_1^{(n)}, \quad
      \{ \bar{Q}, X^{(n)} \} = -P^{(n)} +n J^{(n-1)} + \bar{c}_1^{(n)},
      \nonumber
  \\
  & & \{ S, \bar{X}^{(n)} \} = -P^{(n+1)} - (n-2\ell+1) J^{(n)} + c_2^{(n)}, \label{another_ext}
  \\
  & & \{\bar{S}, X^{(n)} \} = -P^{(n+1)} + (n-2\ell+1) J^{(n)} + \bar{c}_2^{(n)}.
  \nonumber
\end{eqnarray}
The Jacobi identity for $ \langle D, H, P^{(n)} \rangle $ and $  \langle D, C, P^{(n)} \rangle $
gives the relations
\begin{equation}
  n \alpha^{(n-1)} + 2(\ell-n+1) \beta^{(n)} = 0, \quad
  (2\ell-n) \alpha^{(n+1)} - 2(\ell-n-1) \gamma^{(n)} =0,
  \label{rel1}
\end{equation}
respectively.  We also have an another relation from the Jacobi identity for $ \langle C, H, P^{(n)} \rangle:$ 
\begin{equation}
   \alpha^{(n)} + (2\ell-n) \beta^{(n+1)} + n \gamma^{(n-1)}  = 0.
   \label{rel2}
\end{equation}
The super Jacobi identity for $ \langle H, Q, \bar{X}^{(n)} \rangle, $ 
$ \langle H, \bar{Q}, X^{(n)} \rangle, $ $ \langle C, S, \bar{X}^{(n)} \rangle $ and 
$ \langle C, \bar{S}, X^{(n)} \rangle $ provides us further relations on the central elements:
\begin{equation}
  \beta^{(n)}= n c_1^{(n-1)} = n \bar{c}_1^{(n-1)}, \quad 
  \gamma^{(n)} = -(2\ell-n) c_2^{(n)} = -(2\ell-n) \bar{c}_2^{(n)}.
  \label{rel3}
\end{equation}
We see from the relations in (\ref{rel1}) that $ \alpha^{(\ell)} = 0. $ Applying this to (\ref{rel2}) we have 
the relation $ \beta^{(\ell+1)} + \gamma^{(\ell-1)} = 0. $ 
We see from (\ref{rel1}) again that $ \beta^{(n)} (n \neq \ell + 1) $ and $ \gamma^{(n)} (n \neq \ell - 1) $ 
are functions of $ \alpha^{(n)}. $ Thus by setting $ \alpha^{(n)} = 2(\ell-n) a^{(n)} $ one may write as follows:
\[
  \beta^{(n)} = 
    \cases{ -n a^{(n-1)} & $n \neq \ell+1$ \\ -\gamma^{(\ell-1)} & $n= \ell+1$},
    \quad
  \gamma^{(n)} = 
    \cases{ (2\ell-n) a^{(n+1)} & $ n \neq \ell-1 $ \\ \gamma^{(\ell-1)} & $ n = \ell-1 $ }
\]
It follows that 

\[
  c_1^{(n)} = \bar{c}_1^{(n)} = 
   \cases{ -a^{(n)} & $ n \neq \ell $ \\ -\gamma^{(\ell-1)} \over \ell+1 & $ n = \ell $  },
\]
and
\[
  c_2^{(n)} = \bar{c}_2^{(n)} = 
   \cases{ -a^{(n+1)} & $ n \neq \ell-1 $ \\ - \gamma^{(\ell-1)} \over \ell+1 & $ n = \ell-1 $ }.
\]
This means that the central terms in (\ref{another_ext}) are absorbed by the following 
redefinitions of $ P^{(n)}:$
\[
  \cases{
   P^{(n)} + a^{(n)} \to P^{(n)} & $ n \neq \ell $ \\
   P^{(\ell)} + {\gamma^{(\ell-1)} \over \ell+1} \to P^{(\ell)} & $ n = \ell $
  }.
\]
Therefore the central extensions in (\ref{another_ext}) are all trivial.

\end{document}